\documentclass[preprint,authoryear,12pt]{elsarticle}
\usepackage{amssymb}
\usepackage[dvips]{epsfig}
\usepackage{graphicx}
\usepackage{rotate}
\usepackage{color}
\usepackage{natbib}
\usepackage{pifont}
\journal{Astroparticle Physics}
\begin{document}
\def\deg{^{\circ}}
\def\be{\begin{equation}}
\def\ee{\end{equation}}
\def\dl{\Phi^{\mbox{DL}}}
\def\stl{\Phi^{\mbox{SL}}}
\def\sdl{\Phi^{\mbox{SDL}}}
\def\ds{\Phi^{\mbox{DS}}}
\def\ps{\Phi^{\mbox{PS}}}
\def\diffflux{\mbox{GeV}^{-1}\,\mbox{cm}^{-2}\,\mbox{s}^{-1}\,\mbox{sr}^{-1}}
\def\pointflux{\mbox{GeV}^{-1}\,\mbox{cm}^{-2}\,\mbox{s}^{-1}}
\def\diffunits{\mbox{GeV cm}^{-2}\,\mbox{s}^{-1}\,\mbox{sr}^{-1}}
\def\pointunits{\mbox{GeV cm}^{-2}\,\mbox{s}^{-1}}
\def\en{E_{\nu}}
\def\eg{E_{\gamma}}
\def\ep{E_{p}}
\def\epb{\epsilon_{p}^{b}}
\def\enb{\epsilon_{\nu}^{b}}
\def\enbG{\epsilon_{\nu,GeV}^{b}}
\def\enbM{\epsilon_{\nu,MeV}^{b}}
\def\ens{\epsilon_{\nu}^{s}}
\def\ensG{\epsilon_{\nu,GeV}^{s}}
\def\egb{\epsilon_{\gamma}^{b}}
\def\egbM{\epsilon_{\gamma,MeV}^{b}}
\def\exb{\epsilon_{X}^{b}}
\def\g25{\Gamma_{2.5}}
\def\lumi{L_{\gamma}^{52}}
\begin{frontmatter}
\title{The Energy Spectrum of Atmospheric Neutrinos between 2 and
  200 TeV with the AMANDA-II Detector}
\author[Madison]{R.~Abbasi}
\author[Gent]{Y.~Abdou}
\author[RiverFalls]{T.~Abu-Zayyad}
\author[Christchurch]{J.~Adams}
\author[Madison]{J.~A.~Aguilar}
\author[Oxford]{M.~Ahlers}
\author[Madison]{K.~Andeen}
\author[Wuppertal]{J.~Auffenberg}
\author[Bartol]{X.~Bai}
\author[Madison]{M.~Baker}
% ...
\address[Christchurch]{Dept.~of Physics and Astronomy, University of Canterbury, Private Bag 4800, Christchurch, New Zealand}
% ...
\address[Gent]{Dept.~of Subatomic and Radiation Physics, University of Gent, B-9000 Gent, Belgium}
% ...
\address[Madison]{Dept.~of Physics, University of Wisconsin, Madison, WI 53706, USA}
% ...
\address[Bartol]{Bartol Research Institute and Department of Physics and Astronomy, University of Delaware, Newark, DE 19716, USA}
\address[Oxford]{Dept.~of Physics, University of Oxford, 1 Keble Road, Oxford OX1 3NP, UK}
\address[RiverFalls]{Dept.~of Physics, University of Wisconsin, River Falls, WI 54022, USA}
% ...
\address[Wuppertal]{Dept.~of Physics, University of Wuppertal, D-42119 Wuppertal, Germany}
% ...
\fntext[Erlangen]{affiliated with Universit\"at Erlangen-N\"urnberg, Physikalisches Institut, D-91058, Erlangen, Germany}
\fntext[Bari]{also at INFN - Sezione di Bari, Italy}
\fntext[Goddard]{NASA Goddard Space Flight Center, Greenbelt, MD 20771, USA}
\cortext[cor]{Corresponding authors: Kirsten M\"unich, Wolfgang Rhode
  and Julia K.\ Becker. Contact:
julia.becker@rub.de, phone:
+49-234-3223779}
\begin{abstract}
The muon and anti-muon neutrino energy spectrum is determined from 2000-2003
AMANDA telescope data using regularised unfolding.
This is the first measurement of atmospheric
neutrinos in the energy range 2 - 200 TeV. The result is compared to different
atmospheric neutrino models and it is compatible
with the atmospheric neutrinos from pion and kaon decays. No significant 
contribution from charm hadron decays or
extraterrestrial
neutrinos is detected. The capabilities to improve the measurement of the
neutrino
spectrum with the successor
experiment IceCube are discussed.
\end{abstract}
\end{frontmatter}
\parindent=0cm
\parskip=0.2cm
\section{Introduction}
\label{intro}
At energies above $ 0.1$~TeV, about one cosmic ray particle per
square meter per second reaches Earth. At the highest observed energies,
particles reach more than $10^{20}$~eV, which is far above what can be achieved in man-made accelerators.  The origin of these charged
cosmic rays is still being discussed, as their direction is scrambled
by extragalactic and galactic magnetic fields.
One option for identifying the origin of cosmic rays is the observation of secondary particles produced in cosmic ray interactions in the astrophysical plasmas themselves: if a proton $p$ interacts with ambient matter or photon fields $\gamma$, pionic secondaries are produced via the processes
$p+p\rightarrow \pi+X$ and $p+\gamma\rightarrow
\Delta^{+}\rightarrow n+\pi^{+}/p+\pi^{0}$, respectively \citep{ppb2008}.
The charged pions subsequently decay into neutrinos,
$\pi^{\pm}\rightarrow \mu^{\pm}+\nu_{\mu}\rightarrow e^{\pm}+\nu_{e}+\nu_{\mu}+\nu_{\mu}$, where we do not distinguish
between neutrinos and anti-neutrinos.
The resulting neutrino flux usually follows the spectral behaviour of the protons, which is predicted to be close to
$dN/dE \propto E^{-2}$ according to Fermi acceleration (\cite{fermi1949,fermi1954}). The conventional 
atmospheric neutrino spectrum due to pion and kaon decay, on the other hand,
shows a spectral behaviour of approximately $dN/dE\propto E^{-3.7}$
\citep[]{Honda,Volkova,gaisser2001,barr2004,honda_04}. 
An additional component of the atmospheric
neutrino flux comes from the decays of hadrons containing charm and
bottom quarks.  This flux, known as the prompt component is expected
to have a spectrum close to
$dN/dE\propto E^{-2.7}$
\citep[e.g.]{naumov_RQPM_89,Costa,naumov_RQPM_01,martin_GBW,honda_04}. The
prompt atmospheric neutrino flux is lower than the conventional
flux but could start to dominate the total spectrum at energies above
about $100$~TeV.
So far, only the conventional neutrino flux is observed \citep{jess_diffuse}. Measurements at high
energies, i.e.\ above $10-100$~TeV, provide an opportunity to reveal an extraterrestrial or a
charm component. 
 At these energies, the Antarctic Muon And Neutrino
Detector Array
(AMANDA) and its successor IceCube are able to make measurements to
look for deviations from the conventional atmospheric neutrino flux.

The AMANDA-II detector was designed for the detection of neutrinos above
100~GeV. 
It is composed of 677 Optical Modules (OMs), each containing a 8-inch, 14-dynode photomultiplier
 tube (PMT) and a voltage divider for the high voltage. The PMTs are optically coupled to the pressure glass sphere with
a silicon gel and can be operated at a high gain of about $1 \cdot 10^9$. 
The optical modules are attached to 19 vertical strings, instrumenting a cylindrical volume of $0.016$~km$^{3}$ (with a radius of 100 m
 and a height of 500 m), see e.g.\ \cite{ty_neutrino2008}. 
Secondary muons in the ice are produced via the process $\nu_{\mu}+ N\rightarrow\mu+X$ \footnote{Here and throughout the paper, we use
the same notation for particles and antiparticles.}.
The muons produce Cherenkov
 radiation if they travel faster than the speed of light in ice (i.e.\ if the muons travel faster than $v>0.8\cdot c_0$, with $c_0$ as the speed of light in vacuum).  Additional Cherenkov radiation comes from the particles produced
 in  muon interactions, such as bremsstrahlung, direct pair production
 and photonuclear interactions, all dominating at muon energies above 1 TeV.
 At higher energies, the sum of the energy loss due to
  stochastic processes (i.e.\ bremsstrahlung, pair production and nuclear
 interaction) is dominant and increases linearly with the energy. The amount of light detected with the
 optical modules  rises with the muon energy and therefore also with the energy of the
 parent neutrino. Thus the detected light amount can be used to
 determine the primary neutrino energy spectrum.
Neutrino-induced muons can be distinguished from atmospheric muons
by selecting events that traverse the Earth and arrive at the
detector from below the horizon. 
Atmospheric muons cannot reach the detector from those
  directions since they are absorbed
on their way through the Earth. 
In this energy
range
($E< 200$~TeV), neutrino absorption in the Earth is not significant. 
Neutrinos can
traverse the matter without loss and some neutrinos interact close to the detector, so that the products of these
neutrino interactions can be observed. 

AMANDA data from the years 2000 to 2003 are analyzed
to determine the energy spectrum of neutrinos, presenting
for the first time the atmospheric neutrino spectrum in the energy range
$2-200$~TeV. In section \ref{nus:sec}, predictions for atmospheric neutrinos
are reviewed. In section \ref{deconvolutionexplained}, a conceptual overview of the
issues involved in deconvolving a spectrum from observed data are discussed. In section \ref{analysis:sec}, more
details of  the data reduction, simulation and  analysis method 
for the deconvolution of the neutrino spectrum are explained. Section \ref{nn:sec} then describes a 
neural network used
 for the construction of an optimal energy-correlated variable, while section \ref{unfolding:sec} shows how the atmospheric
spectrum is determined by regularised unfolding, and discusses the sources of statistical and systematic
uncertainties that enter the calculation.
Section
\ref{results:sec} summarises the results while section \ref{discussion:sec} discusses
them in the context of other experimental results and flux
predictions. Finally, section \ref{conclusions:sec} gives the conclusions
from this analysis along with an outlook on
the possibilities for IceCube, the successor of the AMANDA
experiment.
\section{Atmospheric neutrinos\label{nus:sec}}
When cosmic rays traverse the Earth's atmosphere, hadronic showers 
are produced by their interactions with the atmosphere. Depending on the energy of the
primary cosmic ray, different secondaries can be produced. Up to
energies of $\sim 100$~TeV, 
the flux is dominated by pion and kaon decays. This flux
is usually referred to as the conventional atmospheric neutrino flux. The expected power-law spectrum of conventional atmospheric
neutrinos is typically one power steeper than the primary cosmic ray spectrum: charged pions and kaons have rest-frame lifetimes
of the order of $10^{-8}$~s. High energy pions and kaons travel far
 enough in the atmosphere that they may interact with atmospheric
 nuclei
 before they decay.
As the lifetime increases with the particles' energy,
high-energy pions and kaons have a higher probability to interact before decaying, which steepens the spectrum by one power. An analytic description of
the neutrino spectrum between $100$~GeV and $5.4\cdot 10^{5}$~GeV is given by \cite{Volkova}:
\be
\frac{dN}{dE_\nu d\Omega}\bigg|_{\nu_\mu}\bigg.(E_\nu,\theta)=
A_{\nu}\cdot \left(\frac{E_{\nu}}{{\rm GeV}}\right)^{-\gamma}\cdot\left[\frac{1}{1+6E_{\nu}/E_{\pi}(\theta)}+\frac{0.213}{1+1.44E_{\nu}/E_{K^{\pm}}(\theta)} \right] \,,
\label{volkova:equ}
\ee
with $A_{\nu}=0.0285\,\diffflux$ and $\gamma=2.69$.
Here, $E_\pi$ and $E_{K^{\pm}}$ 
are energy distribution parameters that depend on the
zenith angle $\theta$ \citep{Volkova}. 
Equation (\ref{volkova:equ}) covers the energy range relevant
for this analysis, and will be used to fit the observed spectrum.

The energy spectrum varies with the zenith angle, as a pion traveling through the atmosphere 
horizontally $(\cos(\theta)=1)$  experiences a smaller density gradient than a pion traversing the atmosphere vertically $(\cos(\theta)=0)$.
Thus, nearly vertical pions have a higher probability of interacting with the atmosphere, which 
reduces the flux compared to the horizontal component. 
Other predictions of conventional atmospheric
neutrino fluxes are given by \cite{gaisser2001,barr2004} and
\cite{honda_04,honda2007}, with uncertainties in the modeling of around 15\% \citep{barr_uncertainties2006}.

The  prompt atmospheric neutrino flux is a second component in the
atmospheric neutrino spectrum and is due to the decay of charm and
bottom hadrons, which contain charm quarks. Since $D^{-}$
and $\Lambda_{c}^{\pm}$ hadrons have lifetimes shorter than
$10^{-12}$~s, they decay before any further interaction with the
ambient matter can take place. Thus, these hadrons produce an
isotropic
neutrino spectrum whose shape is close to the 
primary cosmic ray spectrum, i.e.\ $E^{-2.7}$.

The atmospheric neutrino energy spectrum has so far been measured in
the range $1-10$~TeV
\citep{frejus_spectrum,gonzalez_garcia2006,amanda_le}. 
The flux is found to follow the prediction of the conventional atmospheric
flux within uncertainties. Here, the atmospheric neutrino
spectrum extending up to 200 TeV is presented for the first
time. As
theoretical uncertainties increase towards higher energies, the
experimental investigation of the atmospheric spectrum can lead to
conclusions about particle interactions, possible charm contribution
or a possible extraterrestrial component. 

\section{Determination of the neutrino energy spectrum}
\label{deconvolutionexplained}

Determination of the neutrino energy spectrum in a detector like AMANDA 
is complicated by various factors. Neutrino energies are not 
measured directly but are inferred from measuring the energies of
the interaction products of the neutrinos.
 The three flavors of neutrinos produce different secondary patterns of
particles in a detector, with correspondingly different correlations to 
the primary energy. An electron-neutrino will deposit most of its
energy into an electromagnetic or hadronic cascade, which, limited by the resolution
of the energy reconstruction algorithm, will correlate directly to the
primary energy. A tau-neutrino produces  two cascades of
particles, one from the initial interaction and a second from the
decay of the tau particle. If the energy of the event is sufficiently
small, then both of these may be contained in the detector and thus be
representative of the neutrino energy. 
For the main signature in AMANDA, upward moving muon tracks, the neutrino
energy cannot be directly measured, because of the range of the muon. For
a neutrino that interacts inside the detector, the muon may carry away a 
significant fraction of the neutrino energy which is then not measured. 
For high-energy neutrinos that interact very far away, the  energy deposited
in the detector from the final muon will only correlate weakly with the neutrino
energy. 
This further
 degrades the correlation to the initial energy.
Moreover, whatever energy algorithm is used, its
 limited resolution will further degrade the ability
 of the experiment to measure a particle spectrum.

In practise, one uses knowledge of the neutrino physics and detector
response to infer the parameters of the primary neutrino spectrum. 
Mathematically, this inference is represented by an integral equation, 
\begin{equation} 
p(y) = \int A(y,E) \cdot \Phi(E) dE + b(y)
\end{equation}
where $p(y)$ is the probability of observing an event with a
reconstructed parameter vector $y$, $A(y,E)$ represents the
detector response to initial particles of energy $E$,
 $\Phi(E)$ is the neutrino flux and $b(y)$ is
the non-neutrino background. 
Here, $p(y)$ represents the 
distribution of some parameter related to the energy
of the event, e.g.\ the total deposited energy, or a
direct
estimate of the particle energy.  
Once $p(y)$ is obtained and $A(y,E)$ is determined via detector
calibration and simulation, then there are several ways to infer $\Phi(E)$. The simplest is to take
theoretical
estimates of $\Phi(E)$ and see which one results in an 
expected $p(y)$ that fits the data best. If the theory can be parameterised,
then these parameters can be adjusted until a best fit is found and 
errors on the parameters can be  determined. Examples of this
would be direct fitting of the slope $\gamma$ and normalisation
$C$ of a spectrum of the form $\Phi(E) \sim C\cdot E^{-\gamma}$.

Another method employed in AMANDA is to assume the general
form of a theoretical flux from a calculation and fit for a free
normalisation and deviation from the spectral shape. 
These methods are commonly known as {\em forward folding} - the parameters
of the flux are adjusted and then forward propagated through to an
observable which is compared with data. In principle, the flux can be made
theory independent by choosing a parametric form with a large number of 
parameters.  This leads to a problem where small statistical
fluctuations in the observed data may result in unphysical solutions for the 
spectrum, manifesting as bumps and dips in the spectrum. Methods known as regularisation
are employed, whereby constraints on the smoothness of the solution are 
imposed, to control these effects. For the low parameter methods, regularisation is
built into the solution, for instance by the assumption of a power law spectral form.

 As the number of parameters grows, the computational
time to solve for the best fit set of parameters also increases. 
As an alternative, non-iterative methods, collectively
 known as {\em unfolding methods}, exist for the direct solution of the high parameter
problems. The difficulties of regularisation are still inherent in these
direct methods, just as they are in the interative forward folding  methods. These 
considerations aside, the conceptual 
 basis of  unfolding is to use the  direct inverse
of the matrix $A(y,E)$  to  solve for $\Phi(E) = A(y,E)^{-1} p(y)$. 

 The accuracy of a deconvolution of the spectrum using any of
 the methods improves  on finding an
observable $y$ (which is possibly vector-valued) that is well correlated
to the neutrino energy, manifested by minimising the influence of the 
off-diagonal elements of the matrix $A(y,E)$. There is an irreducible 
component of the off-diagonal elements set by the physics of neutrino
interaction and muon propagation. For instance, the fact that the muon
carries away only a fraction of the neutrino energy and then loses 
energy as it propagates to the detector reduces the correlation of the
muon energy to the original neutrino energy. 
The size of the off-diagonal elements depends on how well correlated
the chosen variables are with the neutrino energy. 
In this work, the energies of the muons are reconstructed by using
a neural network, making use of six energy sensitive variables.  The output 
of the neural network, along with two other energy sensitive parameters, form
the observable $y$ for the unfolding. 

We do not directly unfold the atmospheric neutrino spectrum but unfold an intermediate
spectrum which is a convolution of the actual flux with the probability of the neutrinos
having passed through the Earth  and with the efficiency for detection.
Thus, the intial unfolding returns a neutrino spectrum with the same number of events
as observed in the detector. This is then corrected as a function of energy and angle back to
the true atmospheric spectrum.

\section{The data set and the analysis method\label{analysis:sec}}
For the determination of the neutrino energy spectrum from 
$2-200$~TeV, 807 days effective livetime of data taken by the AMANDA detector between 
the years 2000 and 2003 are used. The selection of neutrino candidates for zenith 
angles $\theta>90^{\circ}$ is presented in \cite{jess_diffuse}. In a final step, 
tracks with $\theta < 100^{\circ}$ are removed to minimise the atmospheric muon contamination 
of the neutrino sample. The final sample contains 2972 neutrinos \citep{kirsten_phd} and includes a background of less than 1\%
misreconstructed atmospheric muons \citep{jess_diffuse}. 

\begin{figure}[h!]
\centering{
\epsfig{file=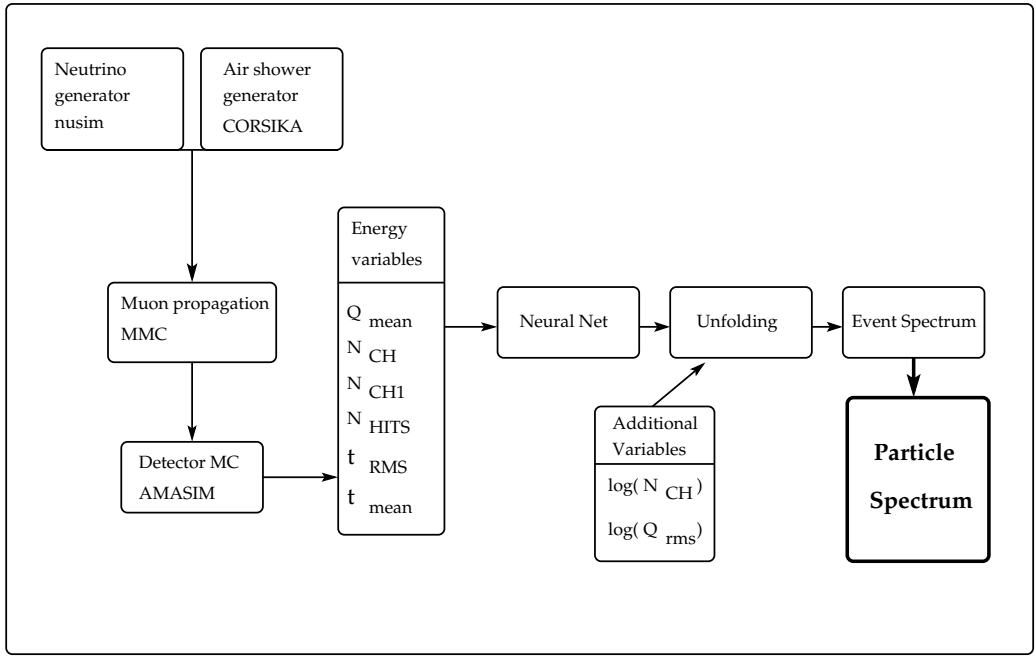,width=\linewidth}
\caption{Scheme of the simulation and analysis chain\label{mc_chain:fig}}
}
\end{figure}
In order to avoid possible biases we perform a blind analysis in the
following sense: The properties of the selected events are initially checked on 10\% of the full 
data set, by comparing this reduced data set to simulations of the atmospheric 
muon background and the conventional atmospheric neutrino flux.
The simulation is done by using the air shower simulation
CORSIKA for the background \citep{corsika} and the neutrino generator {\it nusim}
for the atmospheric contribution \citep{gary_phd1996}, see Fig.\ \ref{mc_chain:fig}. The
neutrino generator takes into account the interaction of the neutrinos
with the Antarctic ice. Then, the tracks of the neutrino-induced muons
are simulated in the muon propagation Monte Carlo {\it mmc}
\citep{chirkin_rhode2004}. Finally, the detector simulation AMASIM
\citep{hundertmark1998} is used to simulate the emitted Cherenkov
light and to emulate the hardware behaviour. 

The same event selection applied to the experimental
data  set are used on the simulated data to identify the
energy-sensitive variables for the analysis. The
simulation data and the 10\% experimental data set are compared to
verify the consistency of the event selection. This is essentially the approach used in a previous AMANDA analysis of the
same upgoing neutrino data set to obtain limits to the extraterrestrial neutrino flux \citep{jess_diffuse}.
 
After having optimised the analysis on the 10\% subset of the data, the full 
data set is used to determine the energy spectrum of the detected up-going 
neutrino-induced events.
These events are expected to pile up 
at energies below 2 TeV. 
Given the high estimated purity of the investigated data
set in terms of neutrinos \citep{jess_diffuse},
           its energy spectrum is expected to be consistent with an
 atmospheric spectrum.
While the theoretical predictions of the shape of this spectrum only
show minor deviations between 
2 TeV and 20 TeV, at energies between 20 TeV and 200 TeV deviations due to the unknown contribution of prompt
 neutrinos - in this case a slight flattening of the spectrum - are possible. Neutrinos of 
extraterrestrial origin are expected to significantly flatten the spectrum at even higher energies. 
 \newline
 
A schematic view of the analysis chain is shown in Fig.\ \ref{mc_chain:fig}. Six
energy-dependent observables are used as input variables for a neural
net in order to produce a combined, optimised energy variable (Section
\ref{nn:sec}). 
The combined variable is taken together with one further observable plus one
of the variables entering the neural net, the latter still containing
a component orthogonal to the NN output. With the three partially independent variables, the neutrino energy spectrum can be determined using regularised unfolding (Section \ref{unfolding:sec}). The resulting spectrum contains the effective  number of events. As consequence of the unfolding, the probability density function of each event is distributed to several bins
of the energy spectrum leading to broken event numbers in the spectrum.
 By relating the simulated neutrino energy spectrum to the effective
 number of events obtained by unfolding the simulated flux, the energy
 spectrum of the effective number of events of the measured data is finally normalised
 to obtain the absolute neutrino energy spectrum.
\section{Selection of energy dependent
 variables\label{nn:sec}}
Due to the limited acceptance and finite resolution of the detector, the reconstruction of the neutrino energy spectrum
can be obtained only with unfolding methods.
For this analysis, the method of regularised
unfolding is used according to the RUN
 algorithm by \cite{blobel_cern,blobel_run}. This
 algorithm allows up to three energy dependent variables as input for the unfolding 
procedure. Therefore, a set of three variables correlated to the neutrino energy is selected.\newline

In a first step, the following seven observables\footnote{Six observables are used as
 first input and an additional one is used at a later stage of the analysis.} which show the best correlation with the generated neutrino energy in simulations, are
 selected from all variables well described by the simulation.

\begin{itemize}
\item $N_{CH1}$: The number of OMs (channels) having detected exactly one photon as signal during the event. Due to the stochastic energy losses of the muons, the number of emitted Cherenkov photons and hit OMs increases
with
 the muon energy.
\item $N_{CH}$: The number of OMs (channels) having detected one or more photons as signal, 
which increases with increasing muon energy. 
\item $N_{HITS}$: The total number of signal photo electrons within the event, which increases with
increasing muon energy. Each OM can contribute to an event by the detection of one ore more photoelectrons
counted in this variable.
\item $t_{mean}$: The average photon arrival time, i.e.\ the sum of all recorded photon arrival times relative to the
 trigger time, divided by the total number of hit optical modules. 
 The higher the neutrino or muon energy, the
 more the mean time is shifted to late arrivals.
\item $t_{RMS}$: The root mean square of the arrival time distribution of the photons, which grows
 with the number of late photons generated i.e. in secondary energy losses.
\item $Q_{mean}$: The sum of all measured charges, in units of photoelectrons,
divided by
 the number of hit OMs. This is equivalent to the mean number of recorded Cherenkov photons and is correlated to the energy.
\item $Q_{RMS}$: The root mean square of the charge distribution of the
photons in each OM,
which grows with the maximal number of photons from secondary energy
losses.
\end{itemize}

Some of these variables are correlated and their number
exceeds the maximum of three allowed as input for RUN. As a consequence of a multitude of tests,
 in a second step, the first six variables are combined through a neural
 network (NN) to 
give one energy dependent variable.
  The NN output is then used together with $\log(N_{CH})$ and $\log(Q_{RMS})$ as the 
three inputs to the RUN algorithm.\newline

\begin{figure}[h!]
\centering{
\epsfig{file=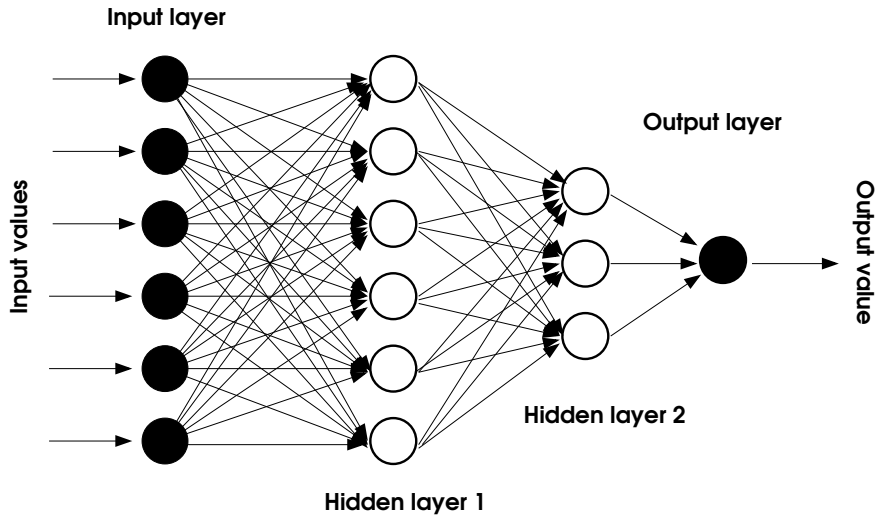,width=12cm}
\caption{Topology of the NN used for the data analysis.\label{nn:fig}}
}
\end{figure}

The neural network is a standard back-propagation Multi-layer Perceptron
 (MLP) with two hidden layers 
used in a 6-6-3-1 feed-forward architecture (Fig.\ \ref{nn:fig}).

\begin{figure}[h!]
\centering{
\epsfig{file=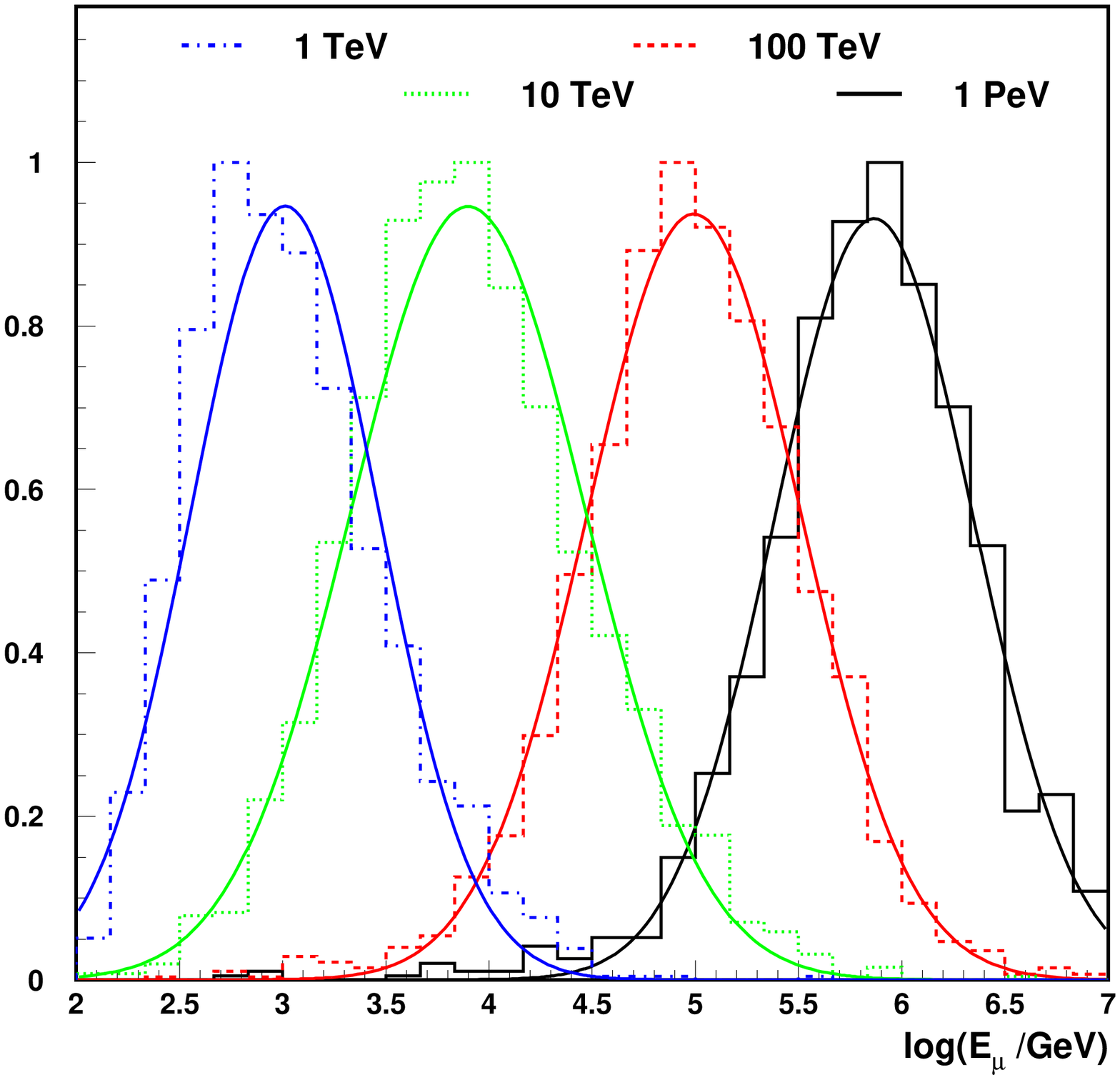,width=9cm}
\caption{Output of NN for the analysis of monoenergetic muons. The muons are simulated with fixed energies of $1,\, 10,\,100$~and~$1000$~TeV.\label{nn_mono:fig}}
}

\end{figure}

The complete simulation chain is used to generate a training set of muon events, with energies uniformly distributed in
      equidistant logarithmic energy bins between 500 GeV and 5 PeV
 and with trajectories uniformly distributed throughout
        a cylindrical volume with 400 m radius around the detector
 center. 
The NN is trained and tested with muon data sets, each containing 100,000 events.
The simulated events are then reconstructed and
        processed in the same way as the experimental data.
The energy resolution of the NN output is
estimated with four test sets of mono-energetic muons, generated in
 the same way as the training set and with muon energies
of 1 TeV, 10 TeV, 100 TeV, and 1 PeV. The resulting performance of the
neural net and the corresponding resolutions are shown in Fig.\ \ref{nn_mono:fig}.  The neural net output can be fitted with a Gaussian
distribution around the logarithm of the expected energy value. The
results for the parameters of the fitted Gaussians are given in Table~\ref{resolution}.

\begin{table}[h]
 \begin{center}
   \begin{tabular}{|l|l|l|}
\hline
     $\log_{10}(E_{\mu}/{\rm GeV})$&$\log_{10}(E_{\rm mean}/{\rm GeV})$&$\log_{10}(\sigma/{\rm GeV})$\\\hline\hline
     3.0&3.03&0.42\\
     4.0&3.92&0.58\\
     5.0&4.99&0.51\\
     6.0&5.86&0.48\\\hline
   \end{tabular}
   \caption{Resolution of the NN output. Here, $E_{\mu}$ is the true input energy,
 $E_{\rm mean}$ is the mean energy of the output and $\sigma$ is the standard 
deviation from $E_{\rm mean}$. All values are given in logarithmic units. \label{resolution}}  
 \end{center}
\end{table}

\begin{figure}[h!]
\centering{%
\epsfig{file=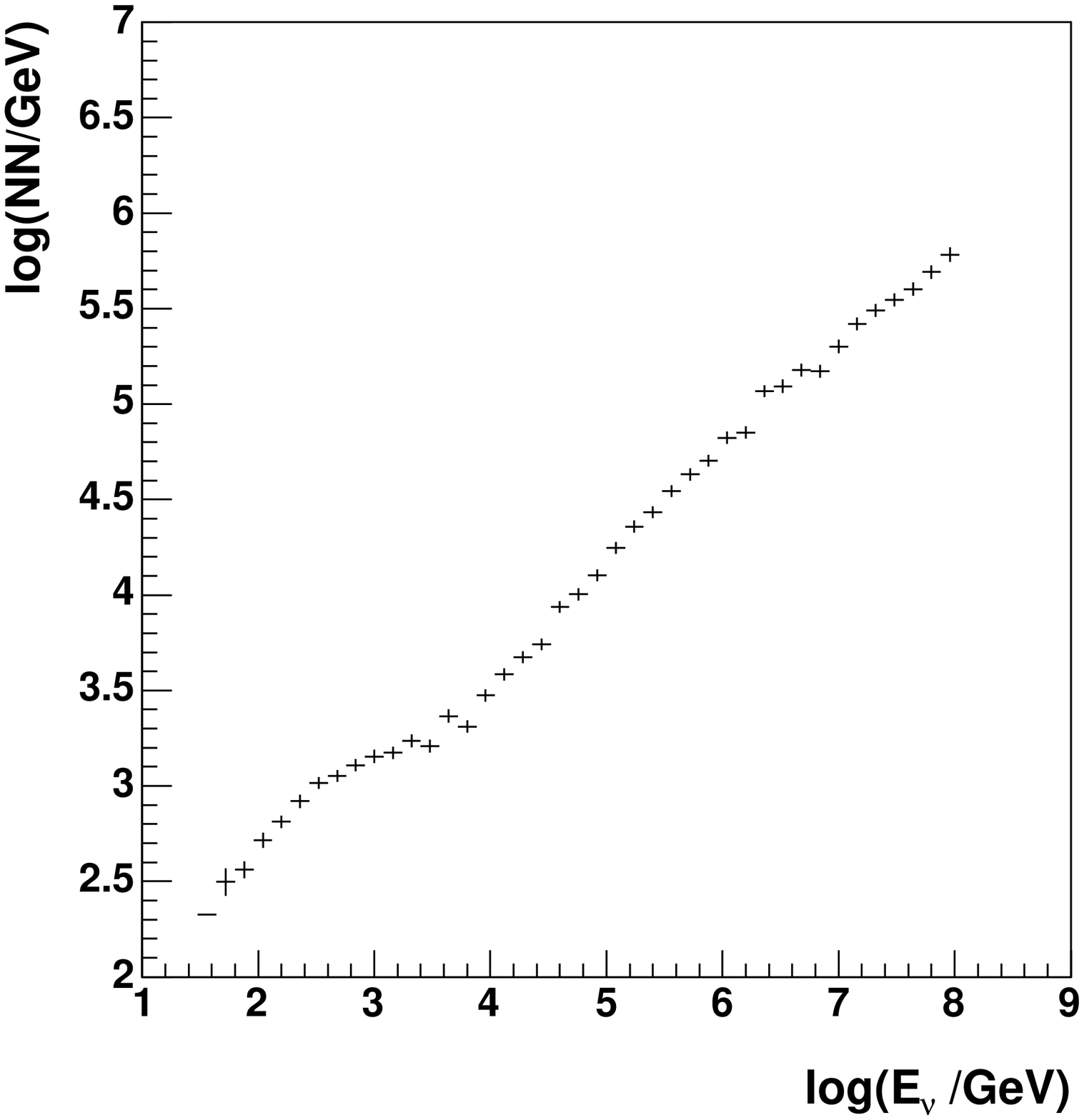,width=8cm}
\caption{
Correlation between NN output and true neutrino energy. Here, the mean value of the NN 
output with its errors as a function of the true neutrino energy is shown. From the NN
output, the neutrino energy can be determined with a standard deviation of about 0.5
order of magnitude.\label{nn_enu:fig}}}
\end{figure}

\begin{figure}[h!]
\centering{
\epsfig{file=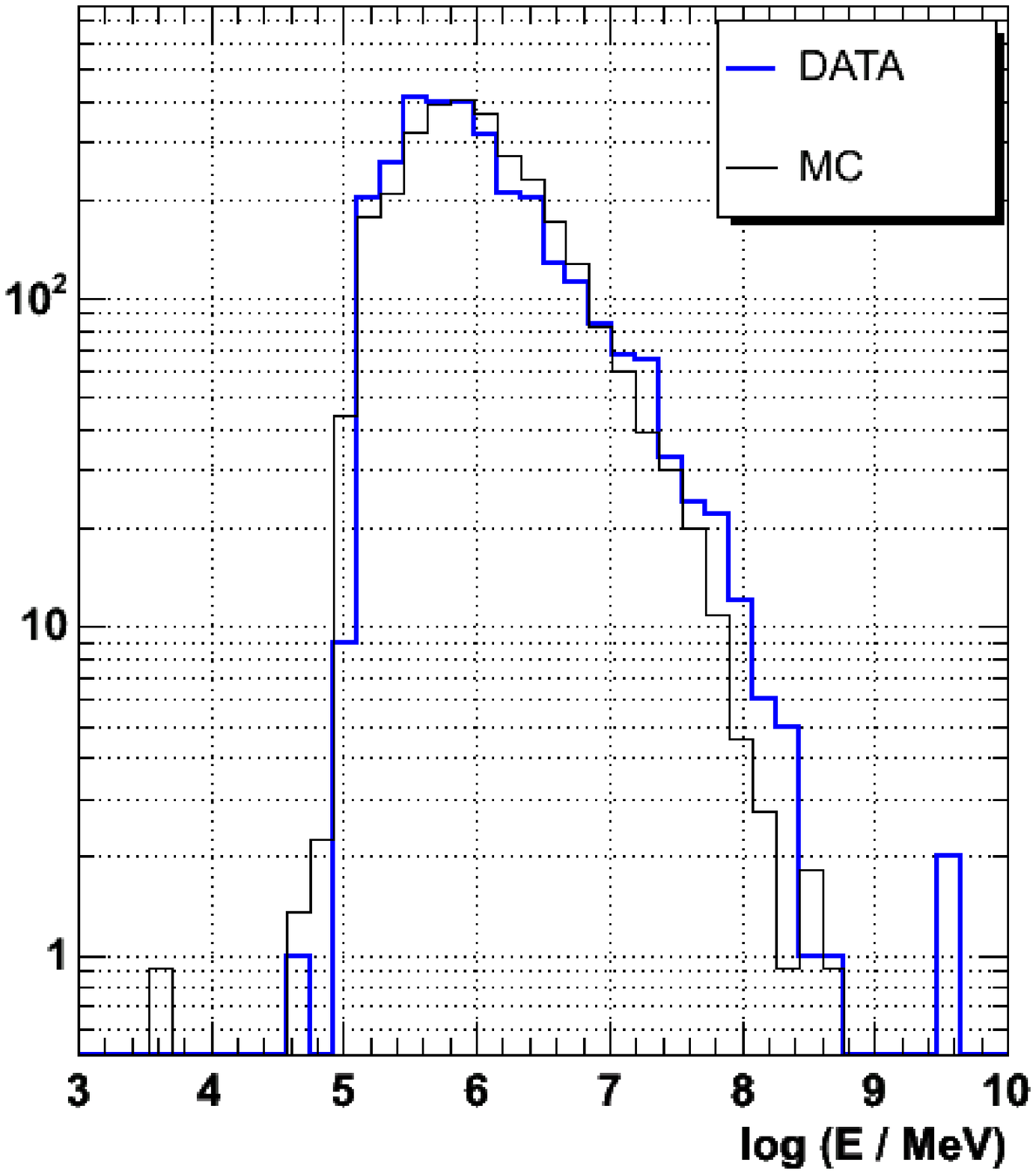,width=9cm}
\caption{Comparison of the neutrino energy NN output for data and a Monte Carlo 
simulation using the parametrisation of atmospheric neutrinos from \citep{honda_04}. \label{nn_out_07:fig}}
}
\end{figure}

Figure \ref{nn_enu:fig} shows how the NN output correlates to the
neutrino energy, and justifies its use as input for the spectrum
 unfolding.
The comparison of the NN output for simulated
 and experimental data (Fig.\ \ref{nn_out_07:fig})
shows good agreement. The agreement in slope depends on the
 neutrino energy spectrum chosen for the simulation. The apparently somewhat steeper
 decrease of the NN output predicted for a simulation according to \cite{honda_04}
 will be consistently visible also in the comparison of the final unfolding result 
(Fig.\ \ref{conventional:fig}) with different flux predictions.


\section{Determination of the energy spectrum \label{unfolding:sec}}

Following the notation introduced in section \ref{deconvolutionexplained}, we 
form the energy sensitive variable $y$ as the vector combination of the
neutral network output, the logarithm of the number of channels fired and the 
logarithm of $Q_{RMS}$.  The probability distribution of this vector variable
is $p(y)$, leading to the need to solve for $\Phi(E)$ in the equation
$p(y) = \int A(y,E) \cdot \Phi(E) dE + b(y)$.

By binning the generated energy distribution and the recorded parameter,
this integral equation can be transformed to a linear matrix
equation 
\begin{equation}
\vec{y}= {\bf A} \cdot \vec{E} + \vec{b}\,.
\end{equation}
The vectors $\vec{y}$, $\vec{E}$ and $\vec{b}$ represent the histograms containing the distribution of
the observable, the sought-after energy spectrum and the distribution of the background. The kernel {\bf A} contains 
the design matrix, describing the
statistical detector properties.
The off-diagonal terms in the kernel arise from the finite resolution of the energy estimators.
Solving this
equation by inversion leads to an ill posed problem because the
transfer matrix {\bf A} necessarily contains off diagonal elements
much smaller than unity, which in turn prevents the calculation of an
{\it a priori} stable solution. To stabilise the solution, proper
assumptions about the curvature of the solution have to be introduced
to cut off insignificant elements of the matrix {\bf A}.\\

In the RUN algorithm \citep{blobel_cern,blobel_run}, the probability distributions used
 to unfold spectra on the basis of given observed parameters are parametrised in the
 form of a superposition of cubic B-splines of fourth order.
 The possible curvature of the solution is controlled by the number of
 degrees of freedom and the number of knots of the spline-superposition.
 If the number of degrees of freedom is too small it
 would damp significant amplitudes and smooth the solution too strongly; 
too many degrees of freedom could enforce unphysical wiggles in the solution.
 The number of knots of the spline is of only little influence on the result if chosen 
much higher than the number of degrees of freedom.
 This procedure is called regularisation and implemented in the unfolding algorithm RUN.

For the given experimental situation, the problem simplifies to the
determination of an approximately linearly decreasing function if, instead of
determining $\Phi(E)$ from $y$, $\log(\Phi(E))$ is calculated from
$\log(y)$. To obtain an optimal parameter combination, extensive
simulation tests of the following form were carried out: first,
the transfer matrix {\bf A} is calculated with an
arbitrary\footnote{It is verified that there are no relevant
  effects for the unfolding results, if training spectra and true spectra do
  not deviate more than $\pm 1$ in the spectral index.} neutrino energy spectrum, a
specific setting of the smoothing regularisation parameter and a
specific binning in $\vec y$ and $\vec E$. 
The used number of degrees of freedom is 5 and the number of knots 26 (see \cite{blobel_cern,blobel_run} for 
a description of the structure).
With this setting of the regularisation parameter, a possible flattening of the spectrum 
is visible in all tested cases.
The 
unfolding result is restricted to be positive
 and the RUN internal histogram used to calculate the
acceptance correction is smoothed.

Using these settings,
different simulated spectra were
unfolded. In total, the unfolding quality is checked with 278,000
Monte Carlo data sets. Each of them contains the unbiased statistical
 equivalent of one year of  AMANDA data, a combination of atmospheric
neutrinos and added signal contributions with an $E^{-2}$  spectrum.
For signal contributions proportional to $E^{-2}$ ranging up to an
contribution of $10^{-6}\,\diffunits$,
1,000 independent unbiased Monte Carlo data sets, each containing one
year of AMANDA data are produced. These Monte Carlo sets were checked to see
if a flattening of the
neutrino energy spectrum towards high energies would be observed if
such a
signal were present.
Also, the statistical 
errors obtained with the algorithm for the given binning  follow a Gaussian 
(or for small effective number of events, a Poissonian) distribution. The  chosen
unfolding parameters
would smooth out spikes on top of the spectrum smaller than the bin width in the 
middle and lower side of the investigated energy range. 

An essential part of the unfolding procedure is the proper estimation and accounting
of the uncertainties, statistical and systematic, that propagate through the 
unfolding to determine the error bars on the spectrum.
In RUN the statistical error is calculated under the assumption that
Poissonian and Gaussian statistics can be applied. 
The same analysis shows that the
distribution of the unfolding results for every bin follows a
Gaussian. It further shows that 
towards high energies a flattening of the spectrum is visible with the chosen
 method if
an extra signal component is present.\\

In addition to the statistical errors,
there are several sources of systematic uncertainty that affect the estimation
of the unfolding matrix and thus propagate through the unfolding to the error bars on 
the physical atmospheric neutrino spectrum.
Estimates of the uncertainties due to the response of the detector are quantified in \cite{5yrs}. Here, we discuss additional sources of uncertainty from the
neutrino cross sections and muon propagation, which both influence the rate and angular
distribution of the detected events.

The neutrino-nucleon DIS cross-section has been measured directly at accelerators up to only $\sim 350$ GeV \citep{ppb2008}. At much higher energies, deep inelastic scattering probes a kinematic region (high $Q^2$ and low Bjorken $x$) where the parton distribution functions have not been directly measured.  The  CC $\nu$ N cross sections used for this paper are calculated as in \citep{giesel,reya}, with details given in \citep{glueck1}. The calculations use the QCD inspired dynamical small $x$ predictions for parton distributions according to the radiative parton model \citep{glueck2} and lead to the conclusion that the systematic uncertainty in the investigated energy range is well
below 10\%. Also, a recent exercise carried out at next-to-leading order using the ZEUS global PDF fits has provided the neutrino cross-section from $10^3$ GeV up to $10^{12}$ GeV with an estimated uncertainty ranging between $\pm 3\%$ and $\pm 14\%$ \citep{cooper_sarkar_2007}.


The systematic errors due to muon energy loss, ice properties and 
effective efficiency of the photomultipliers can be estimated by 
comparing the reconstructed and expected slope of the depth intensity 
relation of atmospheric muons.

This deviation depends on a number of factors relevant for light 
detection, including the muon energy loss, Cherenkov light propagation 
effects and the effective efficiency of the Cherenkov light detection.

 The average range of a muon is approximately $R = 1/b \ln(1 + b/a \cdot 
E)$, where $a$ and $b$ are the effective energy loss parameters for 
ionisation and the sum of the stochastic processes (bremsstrahlung, pair 
production and photonuclear interactions) respectively. They are related 
to the energy loss rate through $dE/dx = a + b\cdot  E$.

 Although the total cross section can be calculated with high precision, 
the spectral averaged energy parameter $b$, in the sense used here 
closely connected to the Cherenkov light  produced by the stochastic 
energy losses, can only be estimated to a precision of a few percent.

 The probability of detecting emitted Cherenkov light depends on both on 
light propagation through the ice and the effective efficiency 
$\epsilon$ of the optical modules.

Comparing the slopes of depth intensity relation of atmospheric muons in 
Monte Carlo and data allows us to set an upper limit on the averaged 
systematic uncertainties. In addition to the factors relevant for the 
muon detection, this deviation also contains effects due to the model 
dependent muon production in the atmosphere. For this analysis, the 
maximal deviation in slope was 10\%.   

This uncertainty in slope transfers directly to the neutrino flux 
calculation. Combining all the independent detector systematics from 
\cite{5yrs} of (8\%) with the cross section (10\%) and the muon 
propagation uncertainties (10\%) gives a total uncertainty in flux of 
16\% which is applied to the statistical error bars from the unfolding.


\section{Energy spectrum of atmospheric neutrinos between $2$~TeV and $200$~TeV\label{results:sec}}

\begin{figure}[h!]
\centering{
\epsfig{file=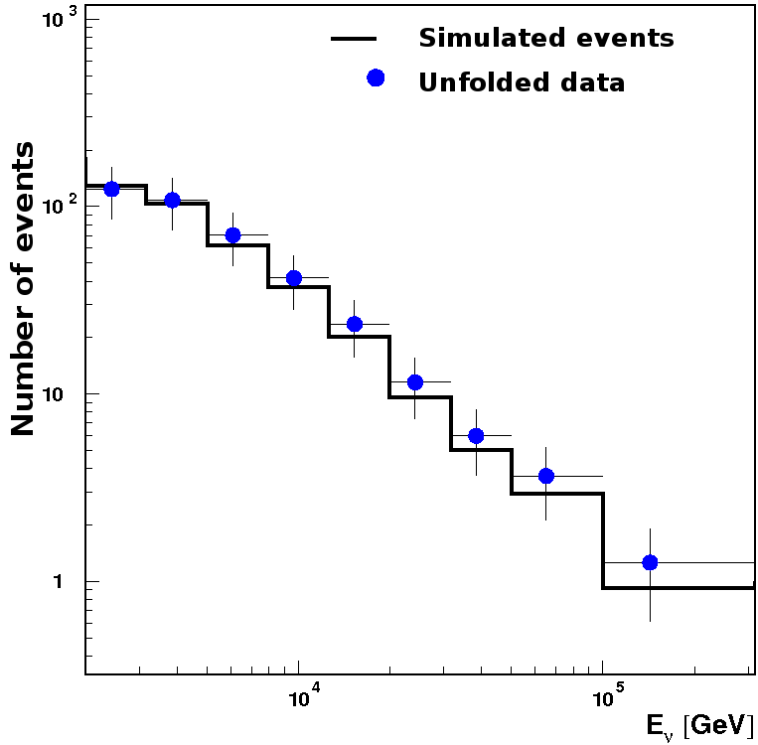,width=10cm}
\caption{ Comparison of the energy distribution of the effective number of events obtained by unfolding the data of the years 2000-2003 (data points) and an event distribution simulated according to \cite{Volkova} (histogram). \label{event_spectrum:fig}}
}
\end{figure}

In a first step, an intermediate energy spectrum, is determined from the data and compared to simulation results in Fig.\ \ref{event_spectrum:fig}. 
This result physically corresponds to
the convolution of the true physical atmospheric
neutrino spectrum and the neutrino survival and detection efficiency, i.e.\ effective area of the detector \citep{jess_diffuse}. The final neutrino spectrum will later be 
found by correcting for the effective area and observation time.
The energy distribution of the effective number of events\footnote{As discussed in section 6, single recorded events contribute in general to
several energy bins of the unfolded energy spectrum. The summed weights 
are called effective event numbers.}
 obtained by unfolding the data of
 the years 2000-2003 and a distribution of events simulated according to \cite{Volkova} are compared. Shown is the number of events
 averaged in the investigated zenith angular range from $100\deg$ to $180\deg$. The error bars include the systematic errors 
of 16\%. For $E> 2$~TeV, the unfolded spectrum agrees with this prediction within the errors.
The corresponding 
number of events for the unfolded spectrum are given in Table \ref{event_spectrum:tab}.
  
\begin{table}
        \centering
                \begin{tabular} {|c|c|}
                \hline
                        $ \log(E_\nu/GeV)$& number of events \\
                        \hline                  
                        $3.3-3.5$&$124$\\
                        $3.5-3.7$&$108$\\
                        $3.7-3.9$&$70.3$\\
                        $3.9-4.1$&$41.6$\\
                        $4.1-4.3$&$23.6$\\
                        $4.3-4.5$&$11.5$\\
                        $4.5-4.7$&$5.96$\\
                        $4.7-5.0$&$3.64$\\
                        $5.0-5.5$&$1.26$\\
                        $5.5-6.0$&$0.00$\\
                        \hline
                \end{tabular}
        \caption{Effective number of events obtained by the unfolding of the neutrino data of the years 2000 to 2003.}
        \label{event_spectrum:tab}
\end{table}

In order to understand and demonstrate the effects of the energy resolution, a further
check is performed. The events in the final simulated data
 set are split up into nine different energy sets based 
on the known true energy of the primary neutrino. Each of these event sets are then 
independently unfolded using the same algorithm as for the full data set. The widths of the
resulting individual spectra give an indication of the energy resolution of the experiment. Each spectrum
is fitted to a Gaussian
\begin{equation}
 F(\log(E_\nu/{\rm GeV}))=A \cdot \exp\left\{-\frac{1}{2}\left(\frac{\log(E_\nu/{\rm GeV})-\log(E_r)}{\sigma}\right)^2\right\}\,.
\end{equation} 
The results of these fits are given in Table \ref{tab:Gauss}. The width of the energy bins denote the energy range to
 which the unfolding and the corresponding errors refer. 
Since the energy resolution obtained with the method described is
between 0.4 and 0.5 in $\log(E_{\nu})$ (see Table \ref{tab:Gauss}), the contents of the bins in the
final spectrum are correlated. 
The
statistical errors obtained by regularised unfolding account for this
fact. 
With the discussed
method, within the investigated energy range a mean energy resolution
of 0.45 in $\log(E_{\nu}/{\rm GeV})$ is reached.

\begin{table}
        \centering
                \begin{tabular} {|c|c|c|c|}
                \hline
                        $ \log(E_\nu/{\rm GeV})$&$A$&$\log(E_r/{\rm GeV}) $&$\sigma $\\
                        \hline
                        $3.5-3.7$&$0.15$&$3.56$&$0.48$\\
                        $3.7-3.9$&$0.16$&$3.76$&$0.46$\\
                        $3.9-4.1$&$0.15$&$4.01$&$0.43$\\
                        $4.1-4.3$&$0.17$&$4.22$&$0.41$\\
                        $4.3-4.5$&$0.18$&$4.41$&$0.43$\\
                        $4.5-4.7$&$0.20$&$4.61$&$0.46$\\
                        $4.7-5.0$&$0.30$&$4.92$&$0.50$\\
                        $5.0-5.5$&$0.44$&$5.39$&$0.49$\\
                        $5.5-6.0$&$0.38$&$5.73$&$0.44$\\
                        \hline
                \end{tabular}
        \caption{Results of the Gaussian fit to simulated data. In the first two columns the energy bins are given, $A_{\nu}$ is the normalisation, $E_r$ the reconstructed mean energy and $\sigma$ the width of the distribution.}
        \label{tab:Gauss}
\end{table}

\begin{figure}[h!]
\centering{
\epsfig{file=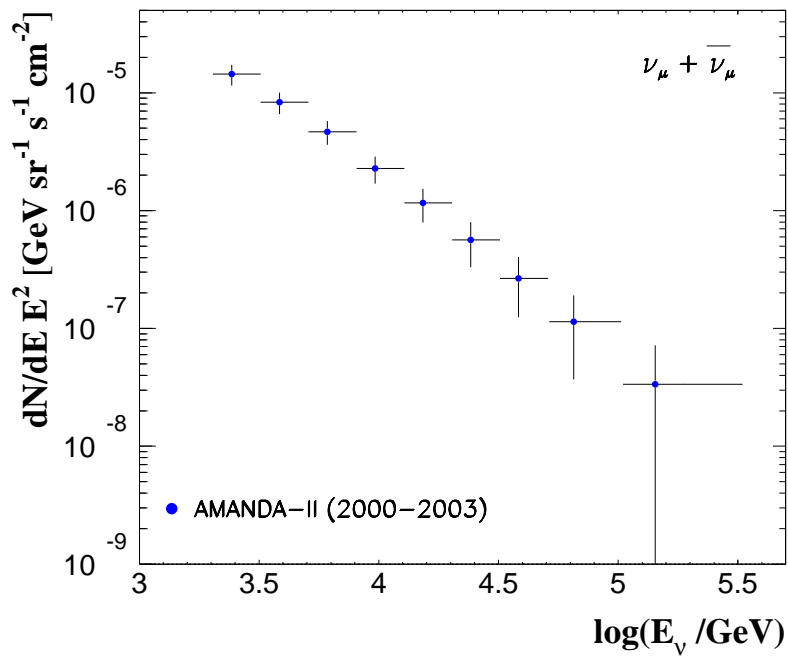,width=12cm}
\caption{The unfolded energy spectrum of muon- and anti-muon neutrinos
  in the atmosphere, measured with AMANDA. 
\label{only:fig}}
}
\end{figure}

In the final step, to obtain the actual energy spectrum of the primary atmospheric neutrinos,
 the ``at the detector'' neutrino spectrum is corrected for the detector efficiency and 
neutrino survival probability. This 
physical energy spectrum of atmospheric muon and anti-muon neutrinos is
presented in Fig.\ \ref{only:fig}. 
Table \ref{nu_energy_spectrum:tab} lists
the values for the measured neutrino spectrum $dN_{\nu}/dE_{\nu}\cdot E_{\nu}^{2}$ for each
energy bin. In the highest-energy bin, the error bars are compatible with a flux
equal to zero. A fit according to Eq.\ (\ref{volkova:equ}), with the normalisation of the
spectrum $A$ and the spectral index $\gamma$ as free parameters yields
\begin{eqnarray}
A_{\nu}&=&(0.022\pm 0.026)\,\diffflux\\
\gamma&=&2.55\pm0.13\,.
\end{eqnarray}
These values are compatible with the theoretical prediction by \cite{Volkova}, $A|_{\rm Volkova}=0.0285\,\diffflux$ and $\gamma|_{\rm Volkova}=2.69$.
The error of $A_{\nu}$ is compatible with zero, since it represents the flux at $1$~GeV, while measurements are performed at above $100$~GeV.

\begin{table}[h!]
\centering{
\begin{tabular}{|l|l|}
\hline
$log(\en/$GeV$)$   &$dN/d\en\cdot {\en}^{2}$\\
& [$10^{-7}$~GeV/s/sr/cm$^2$]\\ \hline\hline
$3.3-3.5$&$140_{-27}^{+26}$\\
$3.5-3.7$&$83_{-16}^{+15}$\\
$3.7-3.9$&$47_{-10}^{+9}$\\
$3.9-4.1$&$23_{-6}^{+5}$\\
$4.1-4.3$&$12_{-3}^{+3}$\\
$4.3-4.5$&$5.6_{-2.4}^{+1.9}$\\
$4.5-4.7$&$2.7_{-1.4}^{+1.0}$\\
$4.7-5.0$&$1.1_{-0.5}^{+0.5}$\\
$5.0-5.5$&$0.34_{-0.34}^{+0.20}$\\
\hline
 \end{tabular}
\caption{The unfolded energy spectrum of muon and anti-muon neutrinos
  in the atmosphere, using AMANDA data from the years 2000-2003.
  The errors give the 68\% C.L. interval on the unfolded flux. 
\label{nu_energy_spectrum:tab}
} }
\end{table}

\section{Discussion of the atmospheric energy spectrum \label{discussion:sec}}
\begin{figure}[h!]
\centering{
\epsfig{file=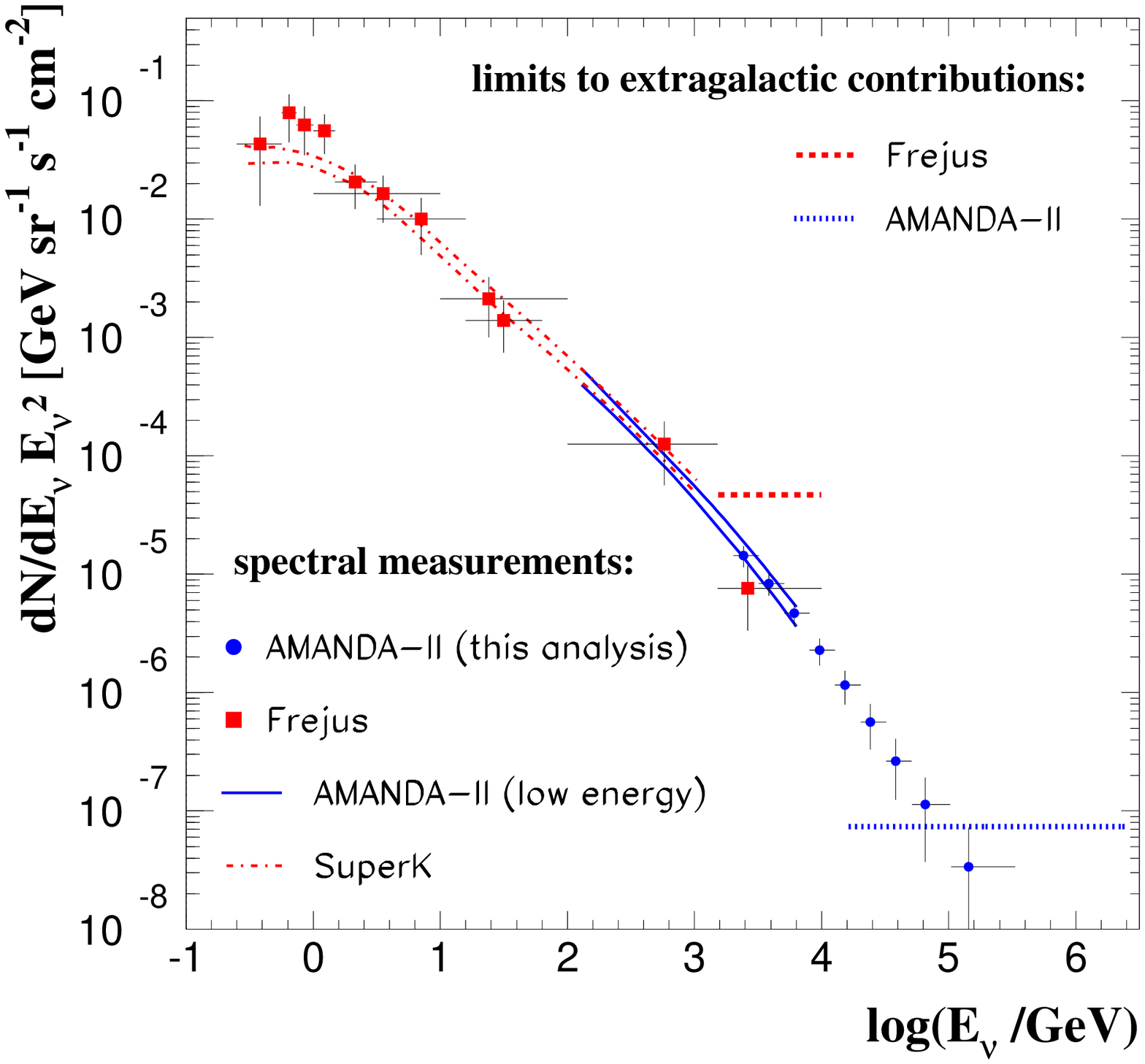,width=\linewidth}
\caption{The unfolded spectrum from this analysis (blue dots)
  compared to other measurements of the atmospheric neutrino
  spectrum: Fr{\'e}jus measurements (red squares) are from \citep{frejus_spectrum}. SuperK results (red
  lines) are presented in \citep{gonzalez_garcia2006}.
 A measurement by AMANDA at
  low energies (blue lines) is shown in \citep{amanda_le}. Limits
  (dashed lines) are from Fr{\'e}jus between $10^{3.2}-10^{4}$~GeV \citep{frejus_limit} and
  from a high-energy analysis with AMANDA using the same data set
  as this analysis \citep{jess_diffuse}.\label{data_comparison:fig}}
}
\end{figure}

Figure ~\ref{data_comparison:fig} compares the unfolded energy spectrum (blue dots) to
previously measured energy spectra. Measurements by the Fr{\'e}jus experiment are shown as red squares \citep{frejus_spectrum}. The red lines 
represent SuperK measurements \citep{gonzalez_garcia2006}.
The latter result is given in the form of a band of possible values indicated in
  this plot as an upper and a lower line. Allowed values (90\%
  confidence level) lie between those two lines. The blue lines at
  higher energies represent AMANDA measurements, based on the same
  data sample, but optimised for low energies \citep{amanda_le}. 
Again, results
  are presented in form of a band, lying between the upper and lower
  line. All measurements are for the sum of neutrinos and antineutrinos. Although the low energy AMANDA analysis is based on the same data set as this analysis, it is
fundamentally different
from the high energy analysis based on an extended regularised unfolding
algorithm discussed here.
The low energy analysis presented in \citep{amanda_le} used the concept of
forward-folding. A set of curves with a limited
number of parameters is used to give an estimate of the input energy
spectrum. In this special case, the
prediction by \cite{gaisser2001,barr2004,barr_homepage} is used with varying normalisation and
spectral index to determine the spectrum.
This method is most sensitive to the median energy of the sample, which is
around 640 GeV and is therefore
not optimal for investigations at high-energies.

These results are the first measurement of the
atmospheric neutrino spectrum at energies up to $200$~TeV. 
Limits to an extraterrestrial neutrino flux with a generic $E^{-2}$
spectrum are shown as dashed lines in Fig.\ \ref{data_comparison:fig}. The Fr{\'e}jus limit is shown at energies between $10^{3.2}$~GeV and
$10^{4}$~GeV. A limit derived from the same AMANDA data set used here
was presented in \cite{jess_diffuse}, confirming that no
significant contribution from extraterrestrial sources at energies
between $10^{4.2}$~GeV and $10^{6.4}$~GeV can be identified at the current sensitivity level.

\begin{figure}[h!]
\centering{
\epsfig{file=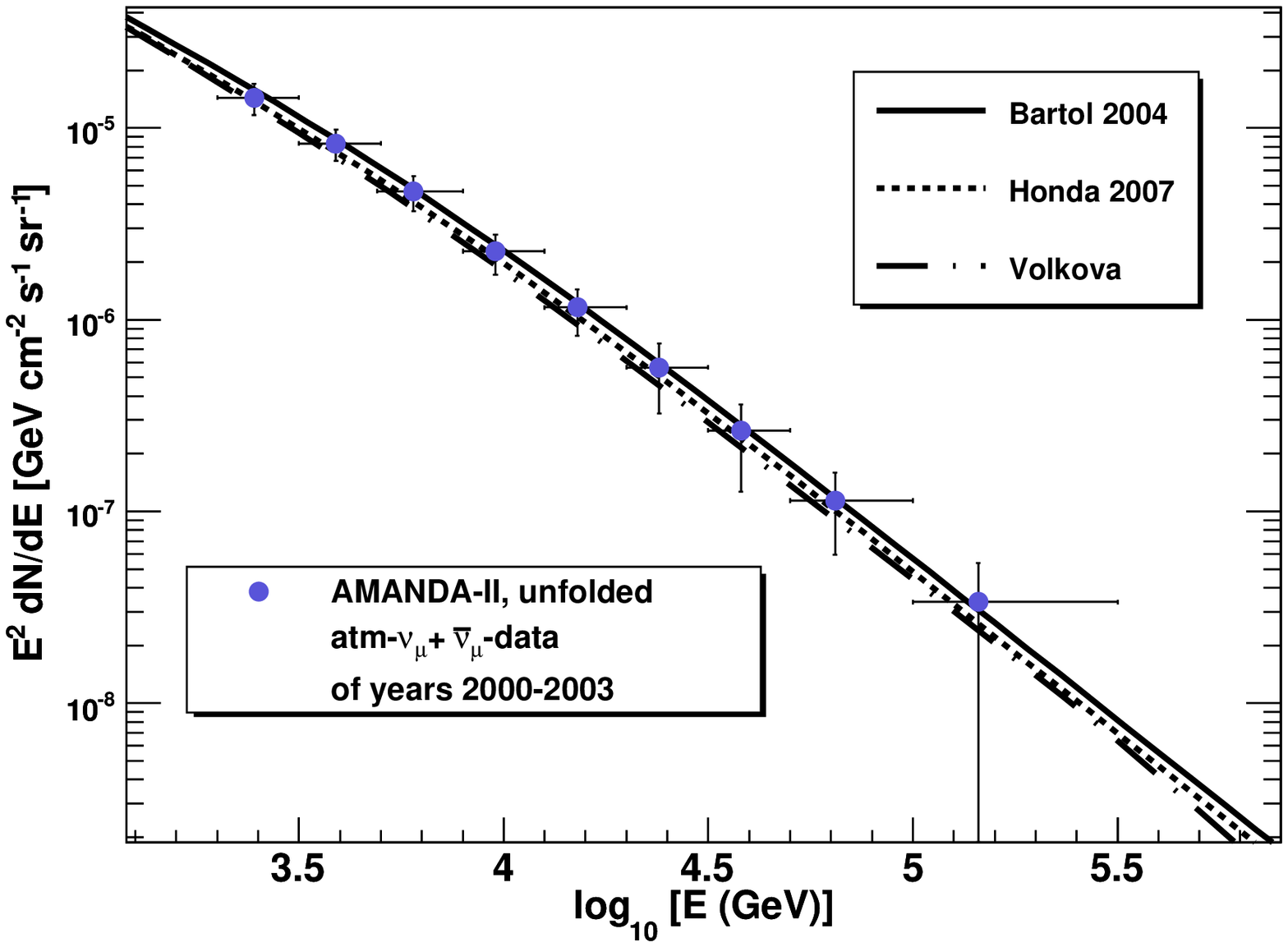,width=\linewidth}
\caption{Comparison of the unfolded energy spectrum to different
  predictions of the atmospheric neutrino flux, resulting from pion
  and kaon decays. The predictions are compatible with the
  measured spectrum within the given errors: \citep{Volkova} (dot-dashed line), \citep{gaisser2001,barr2004,barr_homepage} (solid
  line) and \citep{honda2007} (dashed).  \label{conventional:fig}}
}
\end{figure}

The unfolded neutrino energy spectrum is compared to different predictions of the conventional
neutrino flux in Fig.\ \ref{conventional:fig}. 
As the measured neutrino spectrum includes zenith angles in the range $100^{\circ}<\theta<180^{\circ}$, the predictions are angle-averaged for comparison and the sum of muon and anti-muon neutrinos is used. The conventional atmospheric neutrino flux depends on
parameters which lead to uncertainties in the prediction of the expected flux \citep{barr_uncertainties2006}. We compare the
measured result with different predictions for this flux.
The analytic approximation by \cite{Volkova} is shown as the dot-dashed line. The solid line
represents the Bartol prediction
\citep{gaisser2001,barr2004,barr_homepage}. The flux calculcated by \cite{honda2007} is shown
as the dashed line. The measured spectrum is in good agreement
with all three predictions for conventional neutrinos. 

\begin{figure}[h!]
\centering{
\epsfig{file=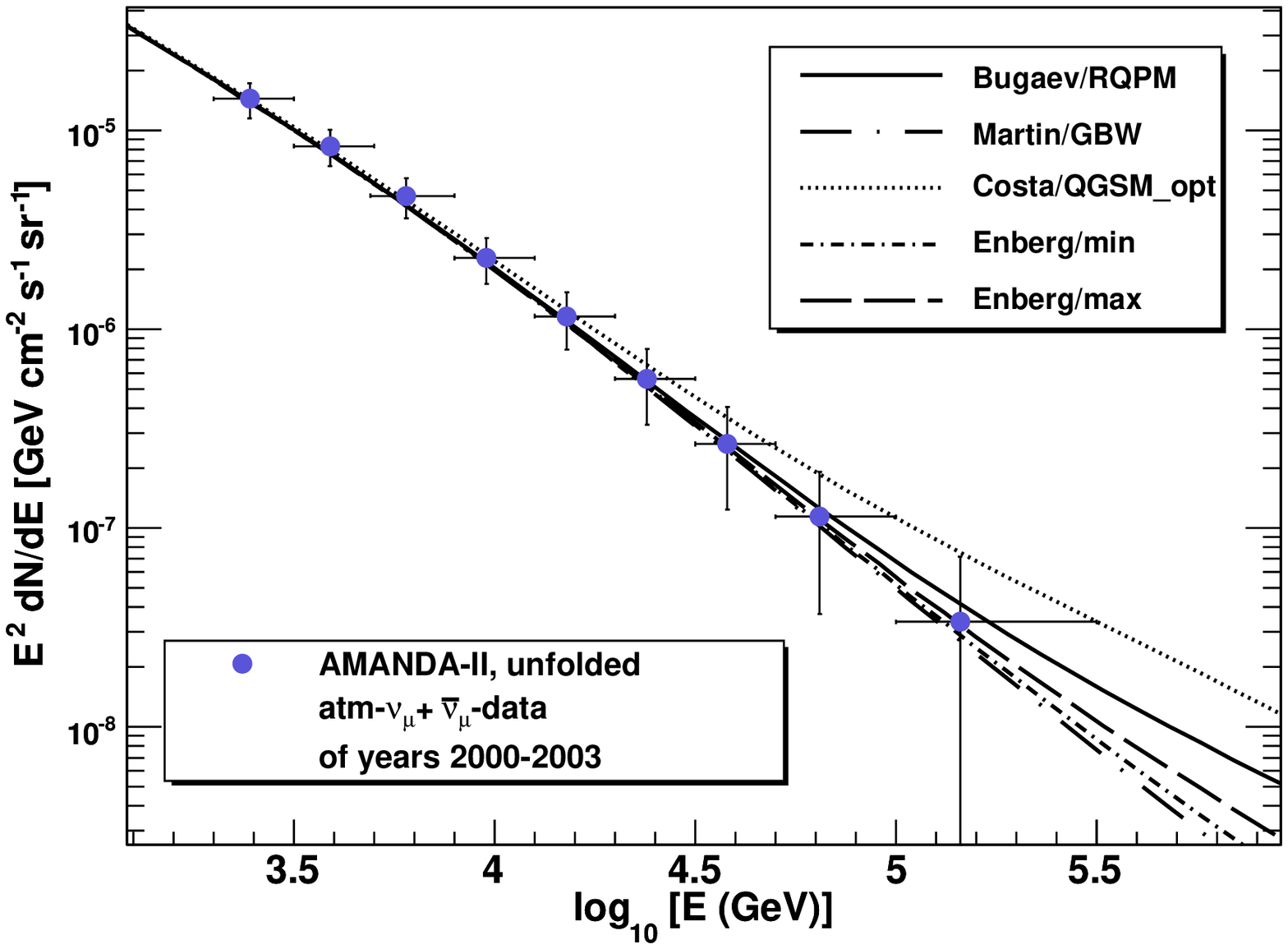,width=\linewidth}
\caption{
The measured neutrino energy spectrum compared to predictions
  of a combination of conventional neutrinos \citep{honda2007} and 
prompt neutrinos. The different prompt models are: \cite{naumov_RQPM_01,naumov_RQPM_89}
  ({\sl Bugaev RQPM}, solid line); \cite{martin_GBW}
  ({\sl Martin GBW}, dot-long-dashed line); \cite{Costa} ({\sl QGSM-opt}, dotted line) and
\cite{enberg2008} ({\sl Enberg/min, Enberg/max}, dot-short-dashed and dashed lines) \label{charm:fig}}
}
\end{figure}

In Fig.\ \ref{charm:fig}, the measured energy spectrum is compared to the prediction of the combined spectrum of conventional and
prompt neutrinos. For the conventional spectrum, the prediction made by \cite{honda_04} is chosen. For the prompt contribution, several 
different models are shown.  The {\bf R}ecombination {\bf Q}uark {\bf P}arton
  {\bf M}odels ({\sl RQPM}) is phenomenology-based and
  non-perturbative, as described by \cite{naumov_RQPM_01}. The shown {\sl QGSM} model is half-empiric, i.e.\ a combination of
theoretical modeling and accelerator data.
 This model uses the {\bf Q}uark {\bf G}luon {\bf S}tring {\bf M}odel based on non-perturbative QCD
  calculations, presented by \cite{Costa} and
  \cite{naumov_RQPM_89}. Shown is the maximum prediction. Further predictions are given by
  \cite{martin_GBW}. A model by \cite{enberg2008} is shown in its
  minimum and maximum configuration. Uncertainties increase
towards higher energies as elaborated in Section \ref{nus:sec}. The
highest prediction ({\sl QGSM opt}, \cite{Costa}) is still compatible with the error bars of the spectrum presented here. Next-generation experiments like IceCube will have a
higher sensitivity to a prompt component.

\section{Conclusions and Outlook\label{conclusions:sec}}
The unfolded muon and anti-muon neutrino energy spectrum is presented for the energy range $2$~TeV and $200$~TeV, constituting the first measurement at such high energies. The spectrum is compatible with predictions of the conventional and prompt atmospheric neutrino spectra. The AMANDA detector was switched off in May 2009 but its more than 60 times larger successor IceCube is currently being built at the same South Pole location. 
As of February 2010, 79 strings have been deployed and completion is
planned within a year, completing an instrumented volume of $1$~km$^{3}$.

\begin{figure}[h!]
\centering{
\epsfig{file=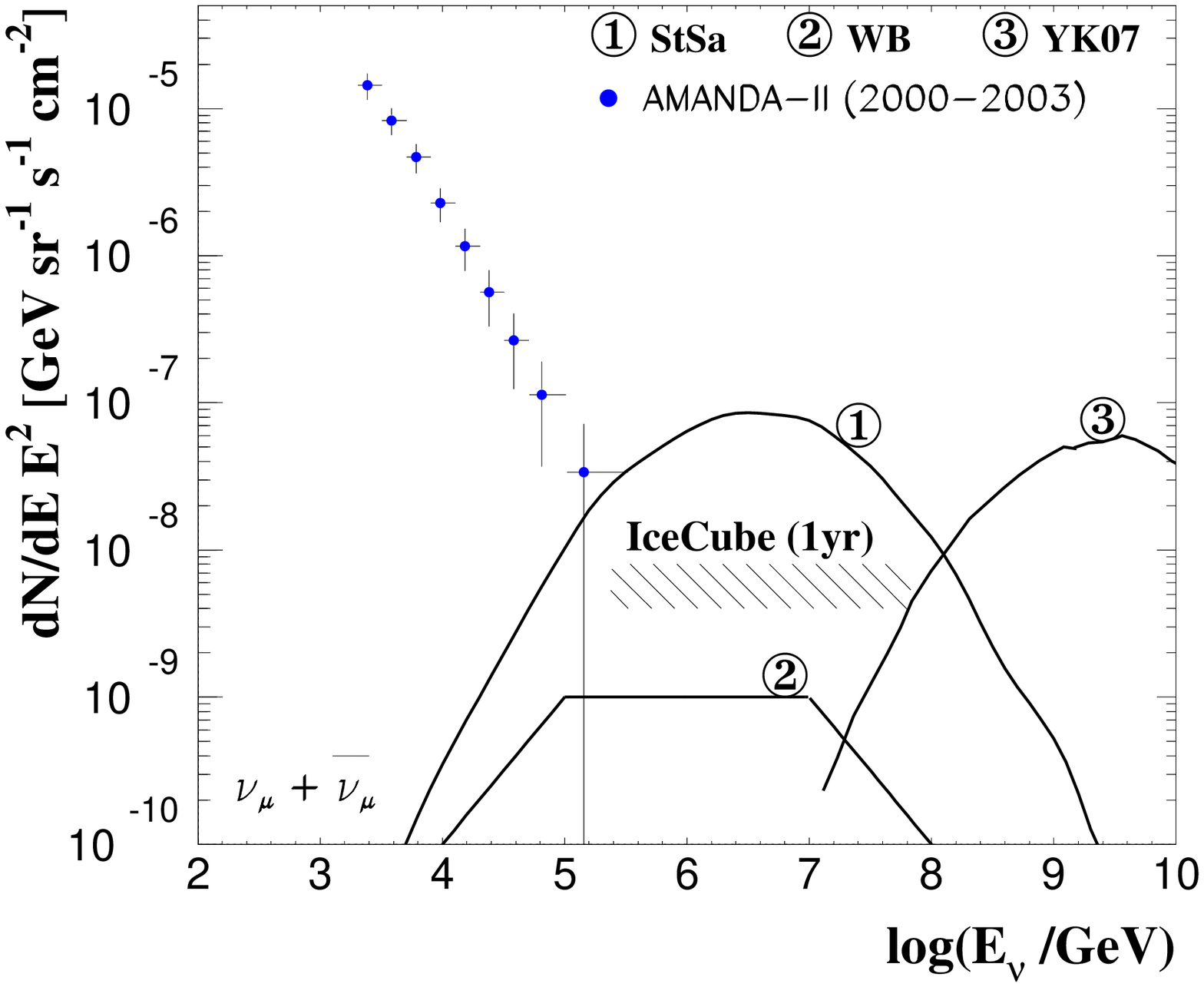,width=\linewidth}
\caption{Measured muon and anti-muon neutrino spectrum and
  predictions of extraterrestrial neutrino fluxes. Neutrinos are expected from e.g.\ Active Galactic Nuclei (e.g.\ \cite{stecker96,stecker_mod}, (1)), Gamma Ray Bursts (e.g.\ \cite{wb97,wb99}, (2)) as well as from the interactions of ultra high-energy cosmic rays with the cosmic microwave background (e.g.\ \cite{yuksel_kistler2007}, (3)). The expected sensitivity of IceCube to an $E_{\nu}^{-2}$ neutrino spectrum is in the range of the hatched area \citep{hoshina2008,francis_sept2008}.\label{extragalactic:fig}}}
\end{figure}

Figure \ref{extragalactic:fig} shows the results of this analysis together with predictions for extraterrestrial neutrino fluxes. Typical neutrino fluxes
from e.g.\ Active Galactic Nuclei or Gamma Ray Bursts are expected to follow a spectrum close to $E_{\nu}^{-2}$, which is much harder than both the conventional ($\sim E_{\nu}^{-3.7}$) and the prompt ($\sim E_{\nu}^{-2.7}$) neutrino flux \citep[e.g.]{halzen_hooper2002,julias_review}. This implies a flattening of the spectrum towards high energies which is much more distinct than for prompt neutrinos. IceCube has the potential to observe this flattening of the spectrum, as its main sensitivity lies in the range $10^{5}-10^{8}$~GeV \citep{hoshina2008,francis_sept2008} and will be able to measure the high-energy neutrino spectrum with higher accuracy and towards higher energies than AMANDA within the first few years of operation.
\clearpage
\subsection*{Acknowledgments}
We acknowledge the support from the following agencies: U.S. National Science Foundation-Office of Polar Program, U.S. National Science Foundation-Physics Division, University of Wisconsin Alumni Research Foundation, U.S. Department of Energy, and National Energy Research Scientific Computing Center, the Louisiana Optical Network Initiative (LONI) grid computing resources; Swedish Research Council, Swedish Polar Research Secretariat, Swedish National Infrastructure for Computing (SNIC), and Knut and Alice Wallenberg Foundation, Sweden; German Ministry for Education and Research (BMBF), Deutsche Forschungsgemeinschaft (DFG), Research Department of Plasmas with Complex Interactions (Bochum), Germany; Fund for Scientific Research (FNRS-FWO), FWO Odysseus programme, Flanders Institute to encourage scientific and technological research in industry (IWT), Belgian Federal Science Policy Office (Belspo); Marsden Fund, New Zealand; Japan Society for Promotion of Science (JSPS); the Swiss National Science Foundation (SNSF), Switzerland; A.\ Kappes and A.\ Gro{\ss} acknowledge support by the EU Marie Curie OIF Program; J. P. Rodrigues acknowledges support by the Capes Foundation, Ministry of Education of Brazil.


\begin{thebibliography}{44}
\expandafter\ifx\csname natexlab\endcsname\relax\def\natexlab#1{#1}\fi
\expandafter\ifx\csname url\endcsname\relax
  \def\url#1{\texttt{#1}}\fi
\expandafter\ifx\csname urlprefix\endcsname\relax\def\urlprefix{URL }\fi

\bibitem[{{Abbasi} et~al.(2009){Abbasi}, {(IceCube Coll.)}, et~al.}]{amanda_le}
{Abbasi}, R., {(IceCube Coll.)}, et~al., 2009. Phys.~Rev.~D 79~(10), 102005.

\bibitem[{{Achterberg} et~al.(2007{\natexlab{a}}){Achterberg}, {(IceCube
  Coll.)}, et~al.}]{jess_diffuse}
{Achterberg}, A., {(IceCube Coll.)}, et~al., 2007{\natexlab{a}}. Phys.~Rev.~D
  76~(4), 042008.

\bibitem[{{Achterberg} et~al.(2007{\natexlab{b}}){Achterberg}, {(IceCube
  Coll.)}, et~al.}]{5yrs}
{Achterberg}, A., {(IceCube Coll.)}, et~al., 2007{\natexlab{b}}. Phys.~Rev.~D
  75~(10), 102001.

\bibitem[{{Amsler} et~al.(2008){Amsler}, {(Particle Data Group)},
  et~al.}]{ppb2008}
{Amsler}, C., {(Particle Data Group)}, et~al., 2008. {Particle Physics Book}.
  Phys.~Lett.~B. 667, 1.

\bibitem[{{Barr} et~al.(2004)}]{barr2004}
{Barr}, G.~D., et~al., 2004. Phys.~Rev.~D 70~(2), 023006.

\bibitem[{{Barr} et~al.(2006)}]{barr_uncertainties2006}
{Barr}, G.~D., et~al., 2006. arXiv:astro-ph/0611266.

\bibitem[{{Barr} et~al.(2009)}]{barr_homepage}
{Barr}, G.~D., et~al., 2009.
  http://www-pnp.physics.ox.ac.uk/$\sim$barr/fluxfiles/.

\bibitem[{{Becker}(2008)}]{julias_review}
{Becker}, J.~K., 2008. Physics Reports 458, 173.

\bibitem[{Blobel(1985)}]{blobel_cern}
Blobel, V., 1985. {Unfolding methods in high-energy physics experiments}. In:
  Proceedings of the 1984 CERN School of Computing. No. CERN 85-09. CERN
  European organization for nuclear research, Geneva, p.~88.

\bibitem[{Blobel(1996)}]{blobel_run}
Blobel, V., 1996. {The RUN Manual - Regularized Unfolding for High-Energy
  Physics Experiments.} Technical Note TN~361, OPAL.

\bibitem[{{Bugaev} et~al.(1989)}]{naumov_RQPM_89}
{Bugaev}, E.~V., et~al., 1989. Nuovo Cim.~C Geophys.~Space Phys.~C 12, 41.

\bibitem[{{Chirkin} and {Rhode}(2004)}]{chirkin_rhode2004}
{Chirkin}, D., {Rhode}, W., 2004. arXiv:hep-ph/0407075.

\bibitem[{{Cooper-Sarkar} and {Sarkar}(2008)}]{cooper_sarkar_2007}
{Cooper-Sarkar}, A., {Sarkar}, S., 2008. J.\ of High Energy Physics 1, 75.

\bibitem[{{Costa}(2001)}]{Costa}
{Costa}, C.~G.~S., 2001. Astropart.~Phys. 16, 193.

\bibitem[{Daum et~al.(1995)}]{frejus_spectrum}
Daum, K., et~al., 1995. Zeitschrift f\"ur Physik C 66, 417.

\bibitem[{{DeYoung} et~al.(2008){DeYoung}, {(IceCube Coll.)},
  et~al.}]{ty_neutrino2008}
{DeYoung}, T., {(IceCube Coll.)}, et~al., 2008. J.\ of Phys.\ Conf.\ Ser.
  136~(2), 022046.

\bibitem[{{Enberg} et~al.(2008){Enberg}, {Reno}, and {Sarcevic}}]{enberg2008}
{Enberg}, R., {Reno}, M.~H., {Sarcevic}, I., 2008. Phys.~Rev.~D 78~(4), 043005.

\bibitem[{{Fermi}(1949)}]{fermi1949}
{Fermi}, E., 1949. Phys.~Rev. 75~(8), 1169.

\bibitem[{{Fermi}(1954)}]{fermi1954}
{Fermi}, E., 1954. Astroph.~Journal 119, 1.

\bibitem[{{Fiorentini} et~al.(2001){Fiorentini}, {Naumov}, and
  {Villante}}]{naumov_RQPM_01}
{Fiorentini}, G., {Naumov}, V.~A., {Villante}, F.~L., 2001. Phys.\ Lett.\ B
  510, 173.

\bibitem[{{Gaisser} et~al.(2001)}]{gaisser2001}
{Gaisser}, T.~K., et~al., 2001. In: Proc. Int. Cosmic Ray Conf. Vol.~5. p.
  1643.

\bibitem[{{Giesel} et~al.(2003){Giesel}, {Jureit}, and {Reya}}]{giesel}
{Giesel}, K., {Jureit}, J., {Reya}, E., 2003. Astropart.~Phys. 20, 335.

\bibitem[{{Gl{\"u}ck} et~al.(1999){Gl{\"u}ck}, {Kretzer}, and {Reya}}]{glueck1}
{Gl{\"u}ck}, M., {Kretzer}, S., {Reya}, E., 1999. Astropart.~Phys. 11, 327.

\bibitem[{{Gl{\"u}ck} et~al.(1998){Gl{\"u}ck}, {Reya}, and {Vogt}}]{glueck2}
{Gl{\"u}ck}, M., {Reya}, E., {Vogt}, A., 1998. European Phys.\ J.\ C 5, 461.

\bibitem[{{Gonzalez-Garcia} et~al.(2006){Gonzalez-Garcia}, {Maltoni}, and
  {Rojo}}]{gonzalez_garcia2006}
{Gonzalez-Garcia}, C., {Maltoni}, M., {Rojo}, J., 2006. Journal of High Energy
  Physics 10, 75.

\bibitem[{{Halzen}(2008)}]{francis_sept2008}
{Halzen}, F., 2008. arXiv:0809.1874.

\bibitem[{{Halzen} and {Hooper}(2002)}]{halzen_hooper2002}
{Halzen}, F., {Hooper}, D., 2002. Reports of Progress in Physics 65, 1025.

\bibitem[{{Heck}(1998)}]{corsika}
{Heck}, D., 1998. {CORSIKA: A Monte Carlo Code to Simulate Extensive Air
  Showers}. Forschungszentrum Karlsruhe Report RZKA 6019.
\newline\urlprefix\url{http://www-ik.fzk.de/corsika/physics\_description/corsi%
ka\_phys.html}

\bibitem[{{Hill}(1996)}]{gary_phd1996}
{Hill}, G., 1996. Ph.D. thesis, University of Adelaide.

\bibitem[{Honda et~al.(1995)}]{Honda}
Honda, M., et~al., 1995. Physical Review D 52, 4985.

\bibitem[{{Honda} et~al.(2004)}]{honda_04}
{Honda}, M., et~al., 2004. Phys.~Rev.~D 70~(4), 043008.
\newline\urlprefix\url{http://link.aps.org/abstract/PRD/v70/e043008}

\bibitem[{{Honda} et~al.(2007)}]{honda2007}
{Honda}, M., et~al., 2007. Physical Review D 75~(4), 043006.

\bibitem[{{Hoshina} et~al.(2008){Hoshina}, {Hodges}, and {Hill}}]{hoshina2008}
{Hoshina}, K., {Hodges}, J., {Hill}, G.~C., 2008. In: Proc. Int. Cosmic Ray
  Conf. Vol.~5. p. 1449.

\bibitem[{{Hundertmark}(1999)}]{hundertmark1998}
{Hundertmark}, S., July 1999. In: Proc. of Simulation and Analysis Methods for
  Large Neutrino Telescopes. DESY Zeuthen, Germany, p. 276,
  {D}ESY-PROC-1999-01.

\bibitem[{{Martin} et~al.(2003){Martin}, {Ryskin}, and {Stasto}}]{martin_GBW}
{Martin}, A.~D., {Ryskin}, M.~G., {Stasto}, A.~M., 2003. Acta Phys.~Pol.~B 34,
  3273.

\bibitem[{M{\"u}nich(2007)}]{kirsten_phd}
M{\"u}nich, K., 2007. Ph.D. thesis, Universit\"at Dortmund.

\bibitem[{{Reya} and {R{\"o}diger}(2005)}]{reya}
{Reya}, E., {R{\"o}diger}, J., 2005. Phys.~Rev.~D 72~(5), 053004.

\bibitem[{{Rhode} et~al.(1996){Rhode}, {(Fr{'e}jus Coll.)},
  et~al.}]{frejus_limit}
{Rhode}, W., {(Fr{'e}jus Coll.)}, et~al., 1996. Astropart.~Phys. 4, 217.

\bibitem[{{Stecker}(2005)}]{stecker_mod}
{Stecker}, F.~W., 2005. Phys.~Rev.~D 72~(10), 107301.

\bibitem[{{Stecker} and {Salamon}(1996)}]{stecker96}
{Stecker}, F.~W., {Salamon}, M.~H., 1996. Space Science Rev. 75, 341.

\bibitem[{Volkova and Zatsepin(1980)}]{Volkova}
Volkova, L.~V., Zatsepin, G.~T., 1980. Soviet Journal of Nuclear Physics 37,
  212.

\bibitem[{{Waxman} and {Bahcall}(1997)}]{wb97}
{Waxman}, E., {Bahcall}, J.~N., 1997. Phys.~Rev.~Lett. 78, 2292.

\bibitem[{{Waxman} and {Bahcall}(1999)}]{wb99}
{Waxman}, E., {Bahcall}, J.~N., 1999. Phys.~Rev.~D 59, 23002.

\bibitem[{{Y{\"u}ksel} and {Kistler}(2007)}]{yuksel_kistler2007}
{Y{\"u}ksel}, H., {Kistler}, M.~D., 2007. Phys.~Rev.~D 75~(8), 083004.

\end{thebibliography}
\end{document}